\documentclass[aps,pra,twocolumn,superscriptaddress,showkeys]{revtex4-1}

\usepackage{graphicx,color}
\usepackage{dcolumn}
\usepackage{rotating}
\usepackage{amssymb}
\usepackage{xspace}
\usepackage{url}
\usepackage[version=3]{mhchem}
\usepackage{tabularx}
\usepackage{threeparttable}

\newcommand{\bra}[1]{\ensuremath{\langle #1 |}\xspace}
\newcommand{\ket}[1]{\ensuremath{| #1 \rangle}\xspace}

\begin{document}
\title{Prospects for high-resolution microwave spectroscopy of methanol in a Stark-deflected molecular beam}
\author{Paul~Jansen$^{a}$, Isabelle~Kleiner$^{b}$, Congsen~Meng$^{a}$, Ronald~M.~Lees$^{c}$, Maurice~H.~M.~Janssen$^{a}$, Wim~Ubachs$^{a}$ and Hendrick~L.~Bethlem$^{a}$\\\vspace{6pt}  $^{a}${\em{Institute for Lasers, Life and Biophotonics, Department of Physics and Astronomy, VU University Amsterdam, De Boelelaan 1081, 1081 HV Amsterdam, The Netherlands}}\\
$^{b}${\em{Laboratoire Interuniversitaire des Syst\`{e}mes Atmosph\'{e}riques (LISA), CNRS UMR 7583 et Universit\'{e}s Paris Diderot et Paris Est, 61 av. G\'{e}n\'{e}ral de Gaulle, 94010 Cr\'{e}teil C\'{e}dex, France}}\\$^{c}${\em{Department of Physics and Centre for Laser, Atomic, and Molecular Sciences, University of New Brunswick, Saint John, New Brunswick E2L 4L5, Canada}}\\}

\begin{abstract}
Recently, the extremely sensitive torsion-rotation transitions in methanol have been used to set a tight constraint on a possible variation of the proton-to-electron mass ratio over cosmological time scales. In order to improve this constraint, laboratory data of increased accuracy will be required. Here, we explore the possibility for performing high-resolution spectroscopy on methanol in a Stark-deflected molecular beam. We have calculated the Stark shift of the lower rotational levels in the ground torsion-vibrational state of \ce{CH3OH} and \ce{CD3OH} molecules, and have used this to simulate trajectories through a typical molecular beam resonance setup. Furthermore, we have determined the efficiency of non-resonant multi-photon ionization of methanol molecules using a femtosecond laser pulse. The described  setup is in principle suited to measure microwave transitions in \ce{CH3OH} at an accuracy below 10$^{-8}$. 
\end{abstract}


\maketitle

\section{Introduction\label{sec:introduction}}
Theories that extend the Standard Model of particle physics have presented scenarios that allow for, or even predict, spatial-temporal variations of the constants of nature \cite{Uzan2003}. Possible variations of the fine structure constant, $\alpha$, representing the strength of the electromagnetic force, or the proton-to-electron mass ratio, $\mu$, a measure of the strength of the strong force, lead to shifts in the spectra of atoms and molecules. Many studies have been devoted to observe these shifts. By comparing metal absorptions in the spectra from distant quasars with the corresponding transitions measured in the laboratory, Webb~\emph{et al.}~\cite{Webb2001} found evidence that suggests that the fine structure constant, $\alpha$, has a smaller value at high redshift. In later work, this variation was interpreted as a spatial variation of $\alpha$~\cite{Webb2011}. In parallel, laboratory experiments on earth are used to probe possible variations in the current epoch. Compared to their astrophysical counterpart, their advantage is their great accuracy, reproducibility and unequivocal interpretation. By comparing transitions in different isotopes of dysprosium, a possible variation of the fine structure constant was found to be $<2.6\times 10^{-15}$/yr~\cite{Cingoz2007}. Whereas atomic spectra are mostly sensitive to variations in $\alpha$, molecular spectra can be used to detect a possible variation of $\mu$. The most stringent independent test of the time variation of $\mu$ in the current epoch is set by comparing vibrational transitions in \ce{SF6} with a cesium fountain, which has resulted in a limit for the variation of $\Delta \mu/\mu$ of 5.6$\times$10$^{-14}$/yr~\cite{Shelkovnikov2008}. Tests of $\mu$-variation on cosmological time scales have been performed by comparing spectra of molecular hydrogen measured in the laboratory with those observed at redshifts $z=2-3$, corresponding to a look-back time of $10-12$\,Gyr, constraining $\Delta\mu/\mu<10^{-5}$~\cite{vanWeerdenburg2011}. The most stringent limit on a variation of $\mu$ in the early universe are set by Bagdonaite \emph{et al.} \cite{Bagdonaite2013} from comparing absorptions by methanol in objects at a redshift of 0.89, corresponding to a look-back time of 7\,Gyr, with laboratory data. The uncertainty in the constraint derived by Bagdonaite \emph{et al.} is dominated by the precision of the astrophysical data. However, when more accurate astrophysical data become available, the error in the laboratory data will become significant. In this paper, we investigate the possibilities to increase the precision of selected microwave transitions in methanol. We focus on the four transitions in \ce{CH3OH} observed by Bagdonaite \emph{et al.}, and two transitions in \ce{CD3OH} that -- provided that the precision is significantly enhanced -- might be used for a laboratory test of the time variation of $\mu$.  

Line centers of methanol transitions in the microwave region are typically obtained from absorption measurements in a gas cell, resulting in (Doppler-limited) measurement uncertainties around 50\,kHz corresponding to a relative uncertainty of $\sim$10$^{-7}$~\cite{Xu2008}. For a limited number of lines higher resolution data was obtained by a pulsed molecular beam Fabry-Perot Fourier-transform microwave spectrometer of the Balle-Flygare type~\cite{Balle1981}, reaching accuracies around 20\,kHz~\cite{Lovas1988}. Using a beam-maser setup, two single methanol transitions were recorded with relative accuracies of $\sim$10$^{-8}$~\cite{Heuvel1973}. So far, this is the only study that was able to (partly) resolve hyper-fine structure in methanol. All these studies are based on detecting absorption or emission of the microwave field. A significantly higher precision seems feasible in a Rabi-type setup using lasers to state-selectively detect the methanol molecules. Unfortunately, so far no suitable state-selective detection scheme for methanol has been demonstrated. The only study that reports the detection of methanol by resonance-enhanced multi-photon ionization (REMPI), involved either the repulsive $3s$ Rydberg state or one of the $3p$ Rydberg state; both resulting in broad unresolved bands~\cite{Philis2007}. 

Here, we explore the possibility for detecting methanol molecules using a femtosecond laser, while relying on inhomogeneous electric fields to separate the different quantum states present in the beam. This paper is organized as follows: In Sec.~\ref{sec:methanol} we discuss the energy level structure of methanol, and review the origin of the large sensitivity coefficients that are found in this molecule. Furthermore, we outline the procedure that was adopted to calculate the Stark interaction for methanol. In Sec.~\ref{sec:simulations} we simulate trajectories of methanol molecules through a typical beam resonance setup, using the derived Stark shifts as input. In Sec.~\ref{sec:detection}, we present measurements that determine the efficiency of ionizing methanol molecules using femtosecond laser pulses. Finally, in Sec.~\ref{sec:allandev}, we use the simulations and measured ion yield to estimate the expected accuracy of the described beam resonance setup.  

\section{Theory\label{sec:methanol}}
\subsection{Torsion-rotation levels in methanol}
Methanol is the simplest representative of the class of alcohol molecules and consists of a hydroxyl (\ce{OH}) group attached to a methyl group (\ce{CH3}). The \ce{CO} bond that connects the two parts of the molecule is flexible, allowing the methyl group to rotate with respect to the hydroxyl group. This rotation is hindered by a threefold potential barrier with minima and maxima that correspond to a staggered and eclipsed configuration of the two groups, respectively. For the lowest energy levels, the internal rotation or torsion is classically forbidden and only occurs due to quantum mechanical tunneling of the hydrogen atoms. In order to account for this additional degree of freedom, the normal asymmetric top Hamiltonian has to be augmented with a term that describe the torsion motion. To simplify the calculation, the coupling between overall and internal rotation is partly eliminated by applying an axis transformation to the coordinates of the Hamiltonian (the so-called ''Rho-Axis Method" or RAM). In the rho-axis method, the full torsion-rotation Hamiltonian for methanol becomes~\cite{Hougen1994}:

\begin{equation}
H_\text{RAM}= H_\text{tors} + H_\text{rot} + H_\text{cd} + H_\text{int}.
\label{eq:Hram}
\end{equation}

\noindent
where $H_\text{tors}$, $H_\text{rot}$, $H_\text{cd}$ and $H_\text{int}$ represent the torsion, overall rotation, centrifugal distortion, and higher-order torsion-rotation interaction terms, respectively. This Hamiltonian is implemented in the {\sc belgi} code~\cite{Hougen1994} that we have used to calculate the level energies of methanol. The current version of the code was modified and improved by Xu \emph{et al.}~\cite{Xu2008} in a number of ways useful for treating the large datasets available for the methanol molecule. Furthermore, the code has been optimized to make it faster and a substantial number of higher order parameters has been added. Using a set of 119 molecular constants for \ce{CH3OH} from Ref.~\cite{Xu2008} and 54 constants for \ce{CD3OH} from Ref.~\cite{Xu1998}, the lower energy levels are found with an accuracy $<100$\,kHz. The Hamiltonian of Eq.~\eqref{eq:Hram} is diagonalized in a two-step process to obtain the torsion-rotation energy levels~\cite{Herbst1984}. In the first step, the torsional Hamiltonian ($H_\text{tors}$) is diagonalized:

\begin{equation}
H_\text{tors}=F\left ( p_\alpha - \rho J_z \right )^2+V\left (\alpha \right ),
\label{eq:Htors}
\end{equation}

\noindent
where $F$ is the internal rotation constant, $\rho$ is the ratio of the moment of inertia of the methyl top relative to the moments of inertia of the molecule as a whole, and $V(\alpha)$ is the internal rotation potential barrier, $p_\alpha$ is the internal rotation angular momentum, $J_z$ is the projection of the global rotation on the $z$ molecular axis and $\alpha$ is the torsional angle. The eigenvalues obtained after this first step are the torsional energies for each torsional level that are characterized by the quantum numbers $K$, $\nu_t$ and $\sigma=0$ (A species) or $\pm 1$ (E species). The $A$ and $E$ symmetry species can be considered as two different molecular species in the same sense as ortho- and para ammonia. The torsional eigenfunctions can be written as linear combinations of the basis wave functions~\cite{Herbst1984}: 

\begin{equation}
\ket{K\nu_t\sigma}=\frac{1}{\sqrt{2\pi}}\ket{K}\sum_{k=-10}^{10}{A_{3k+\sigma}^{K,\nu_t}\exp{\left ( i\left [3k+\sigma \right ]\alpha\right )}},
\label{eq:torseigenfunc}
\end{equation}
where $k$ is an integer. In the second step, the rotational factor $\ket{K}$ in the eigenfunctions from Eq.~\eqref{eq:torseigenfunc} is replaced by the full symmetric top wave function $\ket{JKM}$ to generate the basis set used to diagonalize the remaining terms of the Hamiltonian from Eq.~\eqref{eq:Hram}, i.e. $H_\text{rot}$, $H_\text{cd}$, and $H_\text{int}$.

\begin{figure*}[bt]
\centering
\includegraphics[width=1\textwidth]{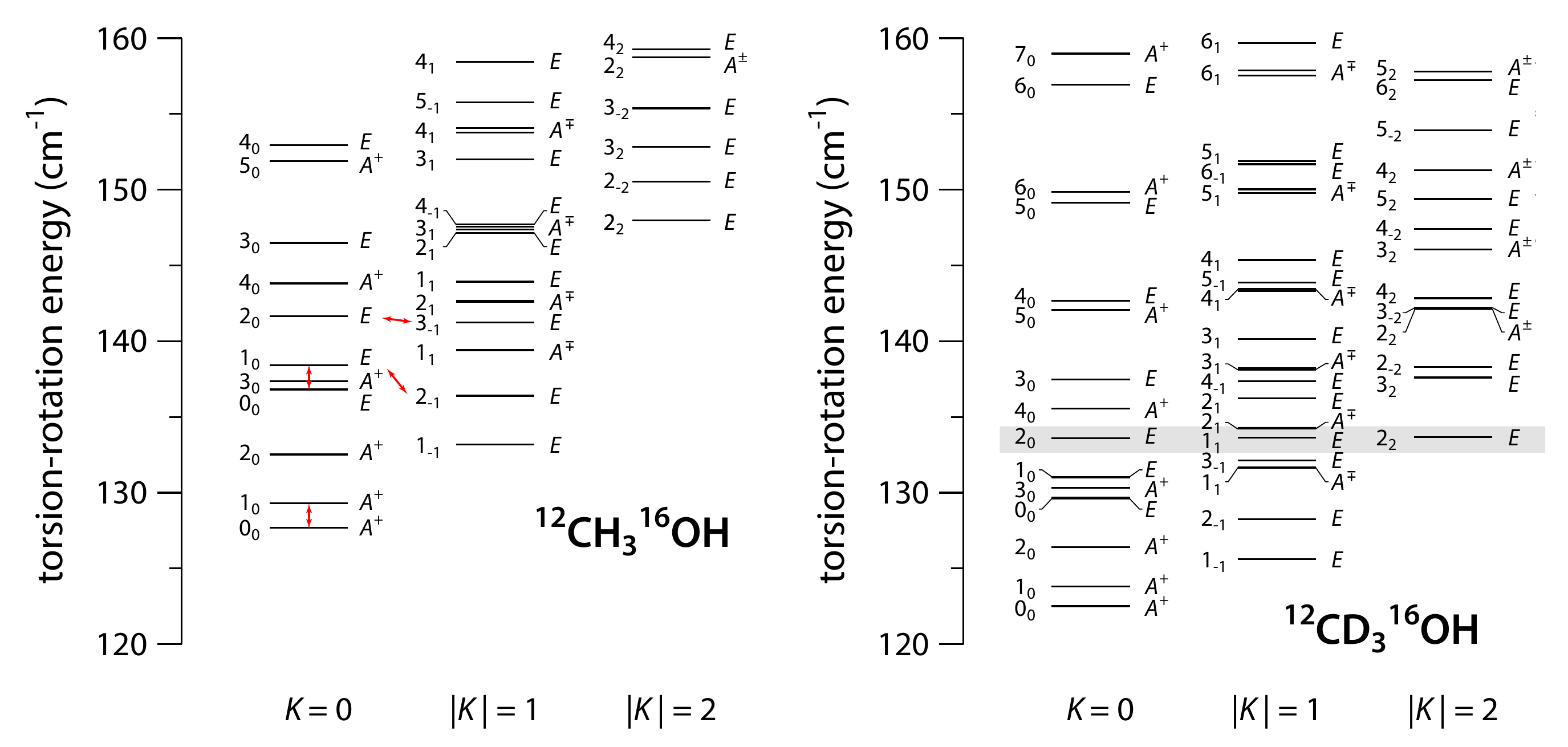}
\caption{Level scheme of the lowest torsion-rotation states in the torsion-vibrational ground state ($\nu_t=0$) of \ce{CH3OH} (left panel) and \ce{CD3OH} (right panel). The energies are calculated at zero electric-field strength using the molecular constants from Refs.~\cite{Xu2008,Xu1998} and are given with respect to the zero point of the torsional well. The levels are labeled by $J_{K}$ (indicated on the left-hand side of each level). For the $A$ levels the so-called parity quantum number (+/-) is also indicated. Arrows in the panel on the left-hand side mark transitions that were used in Bagdonaite \emph{et al.}~\cite{Bagdonaite2013} to constrain $\Delta\mu/\mu$. The shaded area in the panel on the right-hand side highlights the near degeneracies between the $2_0\,E$, $1_1\,E$, and $2_2\,E$ levels in \ce{CD3OH}.
\label{fig:K-ladders}}
\end{figure*}

The lowest energy levels of \ce{CH3OH} and \ce{CD3OH} are calculated at zero electric field using the molecular constants from Refs.~\cite{Xu2008,Xu1998}, and are shown on the left and right-hand side of Fig.~\ref{fig:K-ladders}, respectively. The arrangement of energy levels within a symmetry state resembles that of a prolate symmetric top, with the difference being that every $K$ ladder obtains an additional energy offset due to the $K$ dependent tunneling splitting. As a consequence, certain states in neighboring $K$ ladders may become nearly degenerate. It was shown by Jansen \emph{et al.}~\cite{Jansen2011} and Levshakov \emph{et al.}~\cite{Levshakov2011} that transitions between these nearly degenerate states are very sensitive to possible variations of the proton-to-electron mass ratio, $\mu$. The sensitivity of a transition with a frequency $\nu$ is defined as 

\begin{equation}
\frac{\Delta \nu}{\nu}=K_{\mu}\frac{\Delta \mu}{\mu}.
\label{eq:Kmu}
\end{equation} 

\noindent
The overall rotational energy of the molecule scales with its rotational constants and is thus inversely proportional to the reduced mass of the system. Therefore, pure rotational transitions have a sensitivity coefficient, $K_\mu=-1$. The torsional energy arises from the tunneling effect and -- similar to the inversion splitting in ammonia -- depends exponentially on the effective mass that tunnels. For the normal isotopologue of methanol, the sensitivity coefficient of a purely torsional transition has a value of $K_\mu=-2.5$. Note that, due to symmetry, such transitions are not allowed in methanol. An interesting effect occurs for transitions between different $K$ ladders. In this case, part of the overall rotational energy is converted into internal rotation energy or vice versa. When the energies involved are rather similar -- i.e., when the levels are nearly degenerate -- this results in enhanced sensitivity coefficients. These enhancements occur generally in every internal rotor molecule, but because of a number of favorable properties, the effect is exceptionally large in methanol~\cite{Jansen2011PRA}. 

The arrows in the panel on the left-hand side of Fig.~\ref{fig:K-ladders} mark transitions that have been observed in the study of Bagdonaite \emph{et al.}~\cite{Bagdonaite2013}. The two transitions in the $K=0$ ladder are pure rotational transitions and have a sensitivity coefficient of $K_\mu=-1$. The transitions between the $K=0$ and $|K|=1$ ladder have sensitivities of $K_\mu=-7.4$ and $-33$. The shaded area in the panel on the right-hand side of Fig.~\ref{fig:K-ladders} highlights the near degeneracies present between the $2_0\,E$, $1_1\,E$, and $2_2\,E$ levels in \ce{CD3OH}. The $2_2 \leftrightarrow 1_1\,E$ transition and the $1_{1} \leftrightarrow 2_{0}\,E$ have sensitivity coefficients of 330 and -42, respectively. Thus, if $\mu$ increases, the frequency of the $2_2 \leftrightarrow 1_1\,E$ transition becomes larger while the frequency of the $1_{1} \leftrightarrow 2_{0}\,E$ transition becomes smaller. By comparing these two transitions over a number of years, a possible variation of $\mu$ can be constrained or measured.

\subsection{Stark effect in methanol}
In order to calculate the Stark shift on the energy levels of methanol for different values of the electric field, we have included the Stark Hamiltonian $H_\text{Stark}$ in the second diagonalization step of the {\sc belgi} code. The Stark term is given by

\begin{equation}
H_\text{Stark}=-\vec{\mu}_e\vec{E},
\label{eq:HStark}
\end{equation}

\noindent
where $\vec{\mu}_e$ is the body-fixed electric dipole moment vector and $\vec{E}$ the electric field. In our calculation, $\vec{\mu}_e$ is represented by the body-fixed dipole moments $\mu_a$ and $\mu_b$ along the $a$ and $b$ axes of the RAM frame, respectively. For \ce{CH3OH}, $\mu_a=0.889$\,D and $\mu_b=-1.44$\,D~\cite{Xu2008}, while for \ce{CD3OH}, $\mu_a=0.8340$\,D and $\mu_b=-1.439$\,D~\cite{Jackson1999}. The matrix elements of $H_\text{Stark}$ are taken from Eqs.~(1)--(4) of Kleiner \emph{et al.}~\cite{Kleiner1987}. Since the Stark effect induces nonzero matrix elements $\bra{J}H_\text{Stark}\ket{J\pm 1}$, the Hamiltonian matrix was extended to include those interactions. For a given $J$ value, the Hamiltonian matrix only has off-diagonal blocks involving the nearby $J\pm 1$ states.  Each $J, K$ energy level is split into $2J+1$ components, characterized by $M_J$, the projection of the total angular momentum $J$ along the laboratory axis $Z$. $M_J$ is the only good quantum number in the presence of the electric  field, hence the basis used for calculating the Stark effect for a certain $M_J$ level includes all states with this specific $M_J$.

Our approach was tested by comparing the Stark shifts calculated by the modified version of {\sc belgi} with the results of a perturbation-like calculation using a code written by Lees and coauthors~\cite{Lees1980}. For $J$ levels up to $J=2$, the ratios of the Stark shifts obtained by these two methods are close to unity at 10\,kV/cm and range from 0.7 to 1 at 50\,kV/cm. 

\begin{figure*}[bt]
\centering
\includegraphics[width=1\textwidth]{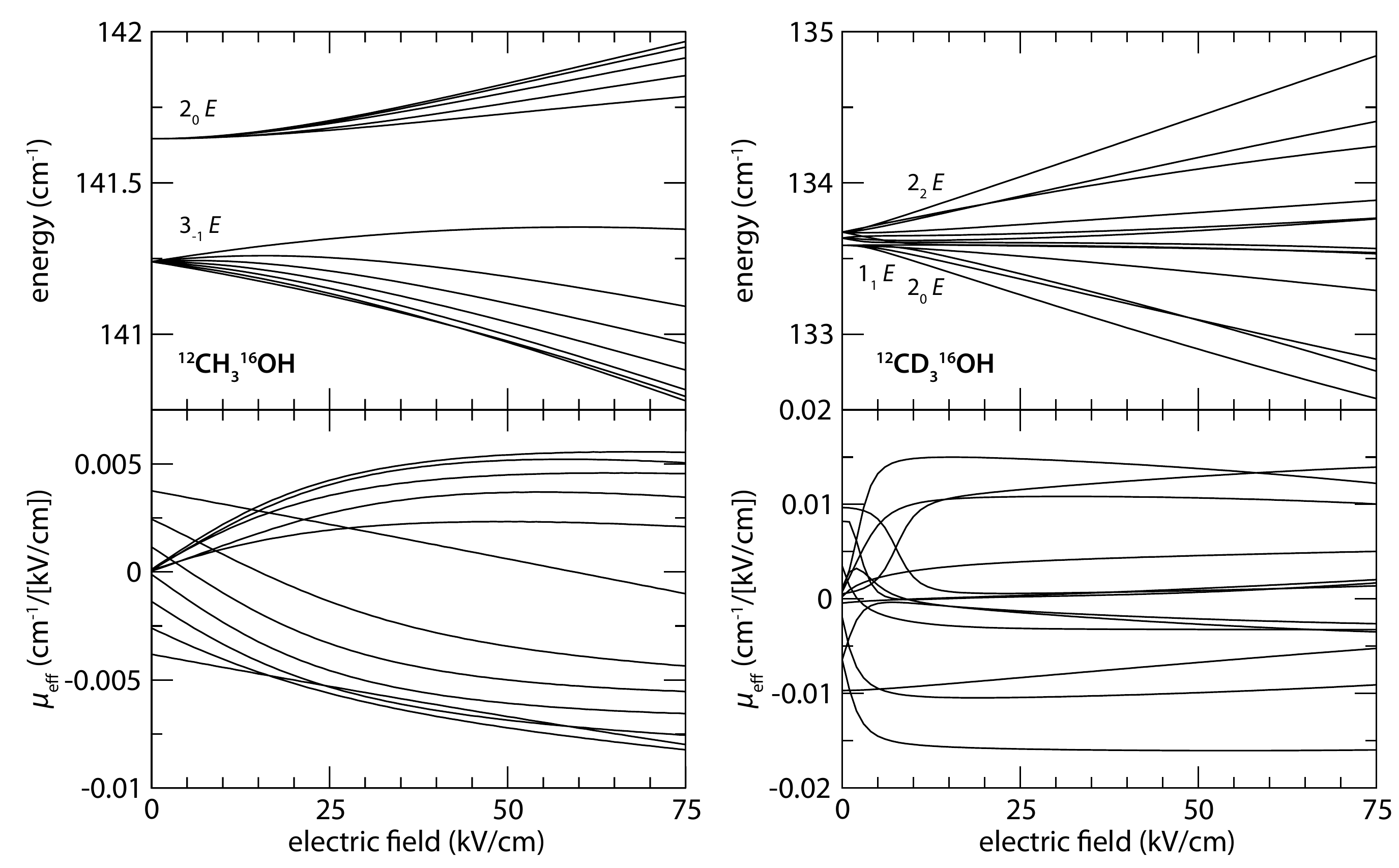}
\caption{Calculated level energies and effective dipole moments of the $3_{-1}\,E$ and $2_{0}\,E$ levels in \ce{CH3OH} (left panel) and $2_0\,E$, $1_1\,E$, and $2_2\,E$ levels in \ce{CD3OH} (right panel) as function of the applied electric field. 
\label{fig:stark_curves}}
\end{figure*}

In the left and right-hand side of Fig.~\ref{fig:stark_curves}, energies of the near-degenerate $3_{-1}\,E$ and $2_0\,E$ levels of \ce{CH3OH} and $2_0\,E$, $1_1\,E$, and $2_2\,E$ levels of \ce{CD3OH}, respectively, are plotted as function of the applied electric-field strength. The lower panels of Fig.~\ref{fig:stark_curves} show the corresponding effective dipole moments $\mu_{\text{eff}}$, defined as

\begin{equation}
\mu_\text{eff} = -\frac{\partial W_\text{Stark}}{\partial |\vec{E}|},
\label{eq:mu_eff}
\end{equation}

\noindent
where $W_\text{Stark}$ is the Stark shift of the quantum state in an electric field of magnitude $|\vec{E}|$. The effective dipole moment is a measure for the orientation of the molecule in a specific state, i.e., the expectation value of the dipole moment in a space-fixed axes system. Note that in the Hamiltonian used, $\ket{J,+K,-M_J}, E$ levels are degenerate with $\ket{J,-K,+M_J}, E$ levels (if $K\neq 0$ and $M_J\neq 0$) and, as these levels have different parity, this degeneracy results in a effective dipole moment that is non-zero at 0\,kV/cm. This non-physical result disappears when high-order couplings are incorporated~\cite{Klemperer1993}. In Table~\ref{tab:mu_eff} the effective dipole moments of a selection of levels are listed. 

\begin{table*}[bth]
\centering
\caption{Effective dipole moments of selected torsion-rotation states of \ce{CH3OH} and \ce{CD3OH}.\label{tab:mu_eff}}
\begingroup
\centering
\scriptsize
\begin{tabular*}{1 \textwidth}{@{\extracolsep{\fill}} p{-0.5cm} D..{0} D..{2} D..{2} c D..{3} D..{3} D..{11} D..{7}}

\hline\hline
\\[-2ex]
& \multicolumn{4}{c}{State} & \multicolumn{1}{c}{Energy (cm$^{-1}$)} & \multicolumn{1}{c}{Energy (cm$^{-1}$)} &\multicolumn{1}{c}{$\mu_\text{eff}$\,(cm$^{-1}$/[kV/cm])} &\multicolumn{1}{c}{$N_i/N$}\\\\[-2ex]\cline{2-5}
\\[-2ex]
& \multicolumn{1}{c}{$J$} & \multicolumn{1}{c}{$K$} & \multicolumn{1}{c}{$M_J$} & Sym & \multicolumn{1}{c}{at 0\,kV/cm} & \multicolumn{1}{c}{at 60\,kV/cm} & \multicolumn{1}{c}{at 60\,kV/cm} & \multicolumn{1}{c}{at 5\,K} \\
\\[-2ex]
\hline
\\[-2ex]
\multicolumn{9}{l}{\ce{CH3OH}}\\
\\[-1.8ex]
&  0	& 0	&  0	& $A^+$ &	127.683 & 127.475  &  6.54\times 10^{-3}   & 1.03\times 10^{-1} \\
&  2	& -1	& -2	& $E$	   &	136.400	& 135.995 &  8.34\times 10^{-3}	 & 2.08\times 10^{-2} \\
&  2	& -1	& -1	& $E$	   &	136.400	& 136.140 &  6.09\times 10^{-3}	  & 2.08\times 10^{-2} \\
&  2	& -1	&  0	& $E$	   &	136.400	& 136.292 &  3.65\times 10^{-3}	  & 2.08\times 10^{-2} \\
&  2	& -1	&  1	& $E$	   &	136.400	& 136.455 &  4.14\times 10^{-4}  & 2.08\times 10^{-2} \\
&  2	& -1	&  2	& $E$	   &	136.400	& 136.586 & -1.10\times 10^{-3}	  & 2.08\times 10^{-2} \\
&  0	&  0	&  0	& $E$	   &	136.805	& 136.701 &  3.00\times 10^{-3}	  & 3.70\times 10^{-2} \\
&  1	&  0	& -1	& $E$	   &	138.419	& 138.425 & -3.45\times 10^{-5} & 2.33\times 10^{-2} \\
&  1	&  0	&  0	& $E$	   &	138.419	& 138.575 & -4.66\times 10^{-3}	  & 2.33\times 10^{-2} \\
&  1	&  0	&  1	& $E$	   &	138.419	& 138.427 & -1.94\times 10^{-4}  & 2.33\times 10^{-2} \\
&  1	& 1	& -1	& $A^+$	&	139.388	& 138.932 &  8.11\times 10^{-3}   & 1.77\times 10^{-3} \\
&  1	&  1	&  0	& $A^+$	&	139.388	& 139.392 & -1.25\times 10^{-4}  & 1.77\times 10^{-3} \\
&  1	&  1	&  1	& $A^+$	&	139.388	& 138.932 &  8.11\times 10^{-3}   & 1.77\times 10^{-3} \\
&  3	& -1	& -3	& $E$	   &	141.240	& 140.908 &  7.20\times 10^{-3}	  & 5.16\times 10^{-3} \\
&  3	& -1	& -2	& $E$	   &	141.240	& 140.899 &  7.65\times 10^{-3}	  & 5.16\times 10^{-3} \\
&  3	& -1	& -1	& $E$	   &	141.240	& 140.927 &  7.18\times 10^{-3}	  & 5.16\times 10^{-3} \\
&  3	& -1	&  0	& $E$	   &	141.240	& 140.978 &  6.32\times 10^{-3}	  & 5.16\times 10^{-3} \\
&  3	& -1	&  1	& $E$	   &	141.240	& 141.051 &  5.28\times 10^{-3}	  & 5.16\times 10^{-3} \\
&  3	& -1	&  2	& $E$	   &	141.240	& 141.154 &  3.87\times 10^{-3}	  & 5.16\times 10^{-3} \\
&  3	& -1	&  3	& $E$	   &	141.240	& 141.354 &  1.52\times 10^{-5} & 5.16\times 10^{-3} \\
&  2	&  0	& -2	& $E$	   &	141.646	& 141.752 & -2.28\times 10^{-3}	  & 9.19\times 10^{-3} \\
&  2	&  0	& -1	& $E$	   &	141.646	& 141.845 & -4.57\times 10^{-3}	  & 9.19\times 10^{-3} \\
&  2	&  0	&  0	& $E$	   &	141.646	& 141.884 & -5.53\times 10^{-3}	  & 9.19\times 10^{-3} \\
&  2	&  0	&  1	& $E$	   &	141.646	& 141.872 & -5.21\times 10^{-3}	  & 9.19\times 10^{-3} \\
&  2	&  0	&  2	& $E$	   &	141.646	& 141.801 & -3.68\times 10^{-3}	  & 9.19\times 10^{-3} \\

\\[-1.8ex]
\multicolumn{9}{l}{\ce{CD3OH}}\\
\\[-1.8ex]
&  2	&  0	& -2	& $E$	   &	133.587	& 133.557 &  1.43\times 10^{-3}   & 9.96\times 10^{-3} \\
&  2	&  0	& -1	& $E$	   &	133.587	& 133.362 &  4.76\times 10^{-3}	  & 9.96\times 10^{-3} \\
&  2	&  0	&  0	& $E$	   &	133.587	& 132.989 &  1.05\times 10^{-2}	  & 9.96\times 10^{-3} \\
&  2	&  0	&  1	& $E$	   &	133.587	& 132.766 &  1.32\times 10^{-2}	  & 9.96\times 10^{-3} \\
&  2	&  0	&  2	& $E$	   &	133.587	& 132.961 &  1.34\times 10^{-2}	  & 9.96\times 10^{-3} \\
&  1	&  1	& -1	& $E$	   &	133.635	& 133.730 & -2.34\times 10^{-3}	  & 4.91\times 10^{-3} \\
&  1	&  1	&  0	& $E$	   &	133.635	& 133.713 & -3.03\times 10^{-3}	  & 4.91\times 10^{-3} \\
&  1	&  1	&  1	& $E$	   &	133.635	& 133.586 &  1.02\times 10^{-3}	  & 4.91\times 10^{-3} \\
&  2	&  2	& -2	& $E$	   &	133.675	& 134.156 & -6.13\times 10^{-3}	  & 4.86\times 10^{-3} \\
&  2	&  2	& -1	& $E$	   &	133.675	& 134.600 & -1.60\times 10^{-2}	  & 4.86\times 10^{-3} \\
&  2	&  2	&  0	& $E$	   &	133.675	& 134.266 & -9.64\times 10^{-3}	  & 4.86\times 10^{-3} \\
&  2	&  2	&  1	& $E$	   &	133.675	& 133.837 & -3.24\times 10^{-3}	  & 4.86\times 10^{-3} \\
&  2	&  2	&  2	& $E$	   &	133.675	& 133.557 &  1.05\times 10^{-3}	  & 4.86\times 10^{-3} \\
\\[-2ex]

\hline\hline
\end{tabular*}
\endgroup
\end{table*}

\section{Trajectory simulations\label{sec:simulations}}
In this section, we use the obtained effective dipole moments to simulate molecular trajectories through a typical beam resonance setup~\cite{RamseyBook}. We consider a molecular beam apparatus that consists of (i) a collimation section; (ii) an electric deflection field for state preparation; (iii) a microwave cavity; (iv) a second electric deflector for state-selection and (v) a detector. In the calculations, we assume a methanol beam with a mean velocity of 800\,m/s. This beam is collimated by two 0.6\,mm diameter skimmers that are separated by 500~mm. The deflection fields used for pre and post state selection consist of a cylindrical electrode and a parabolically shaped electrode to which a voltage difference is applied. The electric field inside such a deflector was analyzed by de~Nijs and Bethlem in terms of a multipole expansion~\cite{deNijs2011}. 
It was shown that the most optimal deflection field is created by choosing a field that only contains a dipole and quadrupole term, the strength of which is represented by $a_{1}$ and $a_{3}$, respectively. In our calculations, we take $a_{1}$=18\,kV and $a_{3}$=3.6\,kV while all other expansion terms are set to 0. Furthermore, $r_{0}$ that characterizes the size of the electrodes is taken as 3\,mm (i.e., the distance between the electrodes is $2r_{0}=6$\,mm). This results in an electric field magnitude of 60~kV/cm on the molecular beam axis. For \ce{CH3OH} the deflectors are assumed to be 200\,mm long, while for \ce{CD3OH} a length of 150\,mm is used. Both deflectors are oriented in the same direction, i.e., a molecule that is deflected upwards in the first electric field, will again be deflected upwards in the second electric field, if it has not undergone a microwave transition. The force on a molecule inside one of the deflection fields is found from:

\begin{equation}
\vec{F} = -\mu_\text{eff}\nabla |\vec{E}|,
\end{equation}

\noindent
where the gradient of the electric field is calculated from an analytical expression derived in de~Nijs and Bethlem~\cite{deNijs2011}. The two deflection fields are separated by 200\,mm. After the last deflection field, the molecules travel 200\,mm further before being ionized in the focus of a femtosecond laser. The total flight path adds up to 1350\,mm for \ce{CH3OH} and 1250\,mm for \ce{CD3OH}. The 200~\,mm free flight between the deflection fields can be used to drive a microwave transition. 

In our simulations, the trajectories of typically $5\times 10^6$ molecules are calculated. The position and velocity spread are sampled randomly from a Gaussian distribution. The initial quantum state is sampled from a Boltzmann distribution that assumes a temperature of either 5 or 10\,K. As mentioned earlier, $A$ and $E$ type methanol should be considered as two different molecular species. In $A$-type methanol, the total spin is $I=\tfrac{3}{2}$, whereas in $E$-type methanol $I=\tfrac{1}{2}$, hence, the nuclear spin degeneracy for $A$-type methanol is twice that of $E$-type methanol. However, this is matched by the presence of both $K>0$ and $K<0$ levels of $E$ symmetry~\cite{Lees1973}. In our calculations, we assume equal numbers of molecules of $A$ and $E$ species for \ce{CH3OH}, whereas the $A/E$ ratio in \ce{CD3OH} is taken to be 11/16~\cite{Lin&Swalen1959}.  

\begin{figure}[btp]
\centering
\includegraphics[width=1\columnwidth]{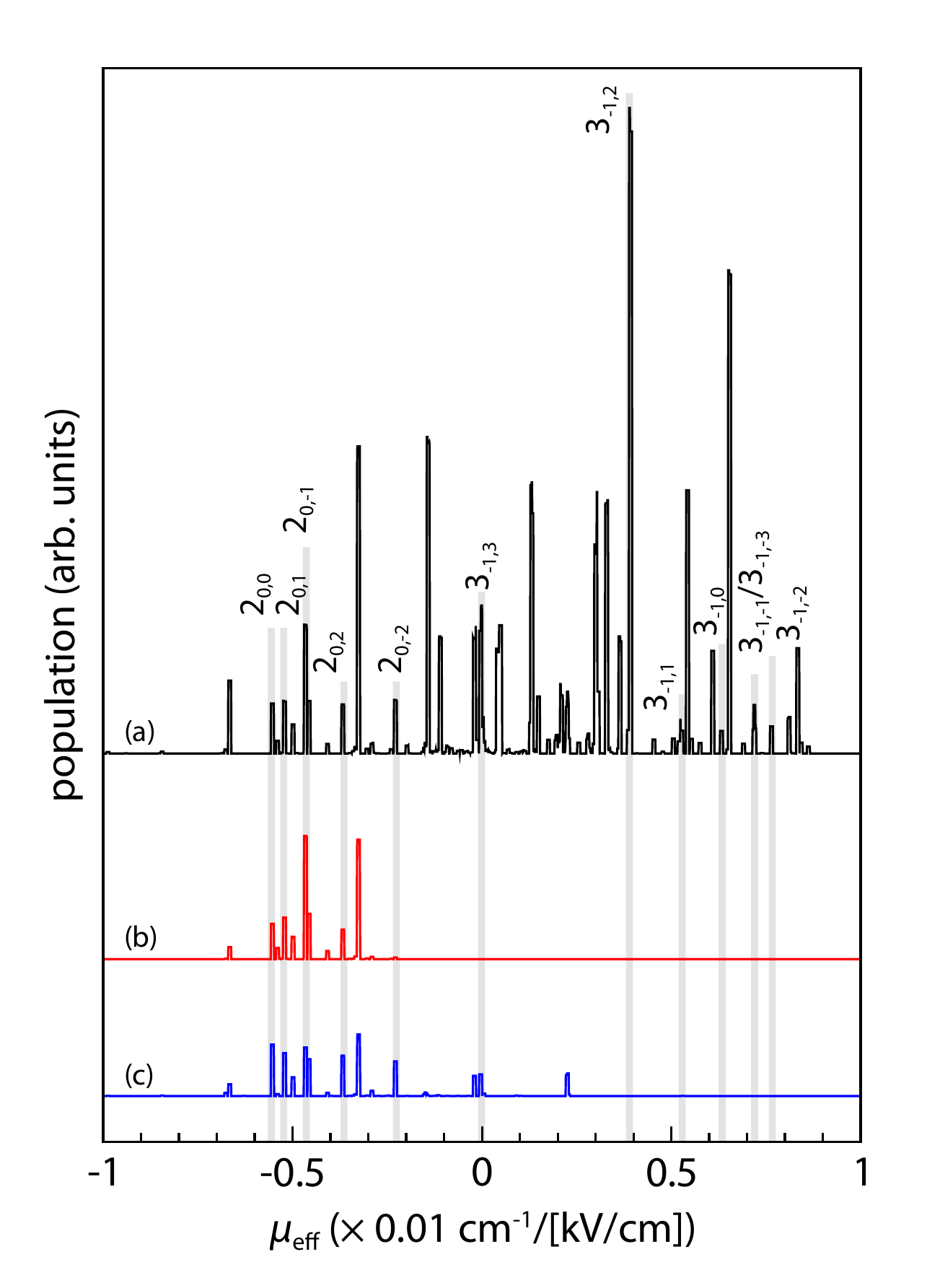}
\caption{Distribution of the effective dipole moments of the methanol molecules in a molecular beam with a temperature of 5\,K. The black curve (a) shows the distribution when the deflection fields are off and the laser focus is situated at the molecular beam axis. The red (b) and blue (c) curve show this distribution when the deflection fields are on and the laser focus is situated 2.2\,mm above the molecular beam axis. For the red curve, $M_J$ is assumed to be preserved in the field-free region, whereas for the blue curve complete $M_J$ scrambling is assumed. The curves are given an offset for clarity.  
\label{fig:mueff}}
\end{figure}

Fig.~\ref{fig:mueff} shows the distribution of the effective dipole moment of the molecules. The black curve (a) shows the distribution when the deflection fields are off and the laser focus is situated at the molecular beam axis. The red (b) curve shows this distribution when the deflection fields are on and the laser focus is situated 2.2\,mm above the molecular beam axis. It is observed that only molecules are detected in states that have an effective dipole moment of $-0.5\,\times$ 10$^{-2}$\,cm$^{-1}$/(kV/cm). With these settings, the resolution of the selector, $\Delta \mu_{\mathrm{eff}}$, is 0.2$\times$ 10$^{-2}$\,cm$^{-1}$/(kV/cm). In these simulations, the $M_J$ state of the molecule is assumed to be preserved -- i.e., it is assumed that a homogeneous electric field is applied between the deflectors to keep the molecules oriented. However, in our experiment the region where the microwave excitation takes place should be completely shielded from external magnetic or electric fields. In this case the different $M_{J}$ levels are degenerate and, in the worst case scenario, the $M_{J}$ distribution after the interaction zone is randomized completely. The results of a simulation that assumes complete de-orientation is shown as the blue curve (c) in Fig.~\ref{fig:mueff}. Clearly, $M_{J}$ scrambling greatly reduces the effectiveness of the state selection. Note that due to hyperfine splittings, the degeneracy of the different levels at zero electric field is lifted. The hyperfine structure has not been taken into account in our simulations as it is largely unknown, in fact, resolving the hyperfine splittings is an important motivation for this study.  

\begin{figure}[bt]
\centering
\includegraphics[width=1\columnwidth]{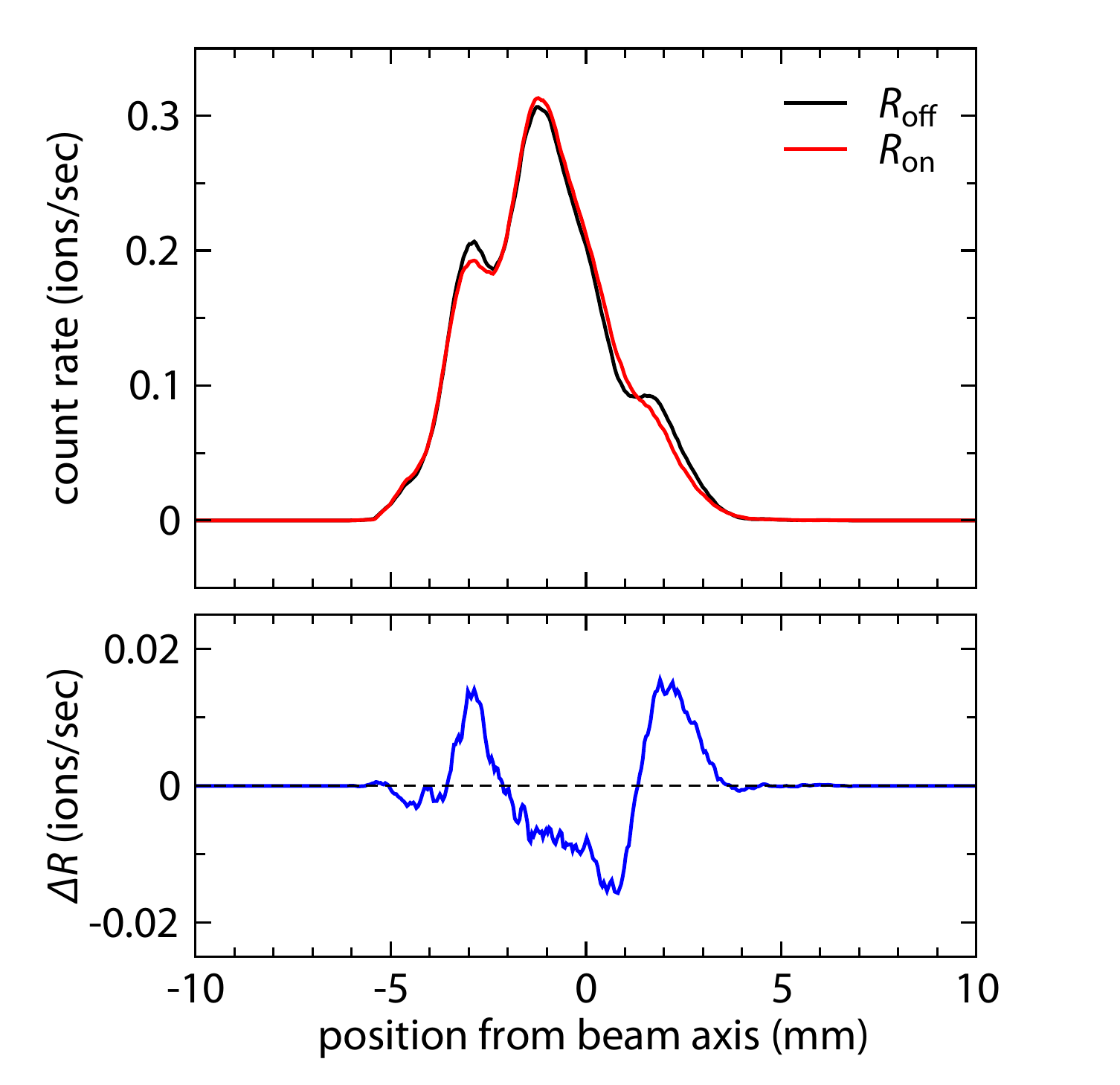}
\caption{\emph{Upper panel:} simulated spatial distribution of the methanol beam along the $y$ axis when the microwave field is off resonance (black curve) and on resonance with the $3_{-1}\rightarrow 2_0\,E$ transition (red curve), assuming a rotational temperature of 5\,K. Both curves are normalized to the number of molecules that are detected at $y=0$ when the deflection fields are turned off. \emph{Lower panel:} difference between the count rate on and off resonance. 
\label{fig:deflection}}
\end{figure}

Fig.~\ref{fig:deflection} shows a simulation of what would be observed if the vertical ($y$) position of the laser focus is scanned, i.e., the spatial distribution of the methanol beam in the direction of field-gradient. The red curve shows the distribution that is observed when the microwave field drives 50\% of the molecules from the $J_{K}=3_{-1}\,E$ to $J_{K}=2_{0}\,E$ level, and \emph{vice versa}. We assume that the microwave field only drives $\Delta M_{J}=0$ transitions. The black curve shows the distribution when no micro-wave field is present. In both cases the $M_{J}$ distribution after the interaction zone is assumed to be randomized completely. Both curves are normalized to the number of molecules that are detected at $y=0$ when the deflection fields are turned off. 
We define the count rate with the microwave field being on or off resonance as $R_\text{on}$ and $R_\text{off}$, respectively. The count rate of the undeflected beam is defined as $R_{0}$. In order to observe if a transition has occurred, we look for a difference between $R_\text{off}$ and $R_\text{on}$, i.e., the difference between the black and red curves in Fig.~\ref{fig:deflection}. This difference is shown in the lower panel of Fig.~\ref{fig:deflection}. If we choose the vertical position of the laser focus to be 2.3~\,mm above the molecular beam axis, $R_\text{off}/R_{0}$= 
0.062 and $R_\text{on}/R_{0}$=0.048. This results in a normalized averaged count rate, $\bar{R}=\tfrac{1}{2}(R_\text{off}+R_\text{on})/R_\text{0}=0.06$, and a difference in count rate, $\Delta R=(R_\text{off}-R_\text{on})/R_\text{0}=0.013$. In Sec.~\ref{sec:allandev}, it is shown that the accuracy scales with $|\Delta R|/\sqrt{\bar{R}}$. For the $3_{-1} \rightarrow 2_0\,E$ transition, this number is equal to 0.056. Table~\ref{tab:transitions} lists $\Delta R$ and $|\Delta R|/\sqrt{\bar{R}}$ for the selected transitions assuming a rotational temperature of 5 and 10\,K. The errors in $\Delta R$ and $|\Delta R|/\sqrt{\bar{R}}$ are estimated to be $10-20\%$ for \ce{CH3OH} and $20-40\%$ for \ce{CD3OH}. For the selected transitions in \ce{CH3OH} $|\Delta R|/\sqrt{\bar{R}}$ is $0.03-0.08$ at 5\,K and slightly less at 10\,K. For the two considered transitions in \ce{CD3OH}, these numbers are even less favorable, which is slightly surprising given that the effective dipole moments of the levels involved are rather different; the different $M_{J}$ states of the $J_{K}=2_{2}\,E$ level all have a positive Stark shift, the $M_{J}$ states of the $J_{K}=2_{0}\,E$ level have a negative Stark effect, whereas the $M_{J}$ states of the $J_{K}=1_{1}\,E$ level have virtually no Stark shift. However, the fact that the effective dipoles of the different $M_{J}$ states within the levels varies considerably, combined with $M_{J}$ scrambling in the interrogation zone, complicates state-selection. Furthermore, the population of the levels involved is rather small.

\begin{table*}[bth]
\centering
\caption{Simulated normalized difference in count rate, $\Delta R=(R_\text{off}-R_\text{on})/R_\text{0}$, and $|\Delta R|/\sqrt{\bar{R}}$ with $\bar{R}=\tfrac{1}{2}(R_\text{off}+R_\text{on})/R_\text{0}$, for selected transitions in methanol at a temperature of 5 and 10\,K. The fourth and seventh columns indicate the $y$ position of the focused laser beam used that results in the maximum ratio of $\Delta R/\sqrt{\bar{R}}$. The third column lists the sensitivity of the transition for a variation of the proton-to-electron-mass ratio, $K_\mu$.\label{tab:transitions}}
\begingroup
\centering
\scriptsize
\begin{tabular*}{1 \textwidth}{@{\extracolsep{\fill}}l D..{3} D..{2} D..{3} D..{2} D..{4} D..{2} D..{2} D..{4}}

\hline\hline
\\[-2ex]

\qquad Transition, $J_K$ & \multicolumn{1}{c}{$\nu$ (MHz)} & \multicolumn{1}{c}{$K_\mu$} & \multicolumn{3}{c}{$T=5$\,K} & \multicolumn{3}{c}{$T=10$\,K}\\\\[-2ex]\cline{4-6}\cline{7-9}
\\[-2ex]
 & & & \multicolumn{1}{c}{$y$\,(mm)}& \multicolumn{1}{c}{$\Delta R$} & \multicolumn{1}{c}{$|\Delta R|/\sqrt{\bar{R}}$}  & \multicolumn{1}{c}{$y$\,(mm)} & \multicolumn{1}{c}{$\Delta R$} & \multicolumn{1}{c}{$|\Delta R|/\sqrt{\bar{R}}$}\\
\\[-2ex]
\hline
\\[-2ex]
\multicolumn{9}{l}{\ce{CH3OH}}  \\
\\[-2ex]
             $\qquad 3_{-1} \rightarrow 2_0\,E$ & 12\,178.587  & -33   &  2.3 & 0.013 &  0.056 &  2.5 & 0.017 &  0.063 \\
             $\qquad 0_0 \rightarrow 1_1\,A^+$  & 48\,372.460  & -1.00 & -3.5 & 0.028 &  0.078 & -3.7 & 0.012 &  0.045 \\
             $\qquad 0_0 \rightarrow 1_0 \,E$   & 48\,376.887  & -1.00 &  2.1 & 0.008 &  0.028 & -1.7 & 0.012 &  0.026 \\
             $\qquad 2_{-1} \rightarrow 1_0\,E$ & 60\,531.477  & -7.4  &  1.6 & 0.012 &  0.041 & -1.2 &-0.016 &  0.029 \\
\\[-2ex]
\multicolumn{9}{l}{\ce{CD3OH}}  \\
\\[-2ex]
             $\qquad 1_1 \rightarrow 2_2\,E$    &  1\,202.296  & 330   &  2.1 & -0.003 &  0.018 & -4.2 &-0.002 &  0.017 \\
             $\qquad 2_0 \rightarrow 1_1\,E$    &  1\,424.219  & -42   & -1.7 & 0.018 &  0.033 & -3.4 & 0.004 &  0.023 \\

\\[-2ex]
\hline\hline
\end{tabular*}
\endgroup
\end{table*}

\section{Non-resonant ionization of methanol using femtosecond laser pulses\label{sec:detection}}

In order to estimate the efficiency of non-resonant multi-photon ionization of methanol by a femtosecond laser, we have performed test measurements in an existing molecular beam machine described elsewhere~\cite{WimR}. A supersonic molecular beam is prepared by expanding a mixture of $\sim$0.5\% \ce{CH3OH} in argon through a 200\,$\mu$m diameter pulsed piezo nozzle~\cite{Irimia2009} operating at a repetition frequency of 10--1000\,Hz. The pressure behind the nozzle was kept below 1 Bar in order to prevent cluster formation. The molecular beam passes a 1.5\,mm and a 1.0\,mm skimmer, before entering the detection region where it is intersected at right angle with a focused ($f=500$\,mm) femtosecond laser beam. The total flight path from the nozzle to the detection zone adds up to 160\,mm. The produced ions are accelerated towards a position sensitive microchannel-plate detector mounted in front of a fast phosphor screen where they are counted. The femtosecond pulses are generated by a commercial amplified regen laser system (Spectra Physics Spitfire) that produces a 1\,kHz pulse train and is tunable around 800\,nm with an output
energy of about 800\,$\mu$J and a duration of 120~fs. The fundamental 800\,nm light is doubled in a BBO crystal to obtain a 400\,nm laser pulse with an energy of about 100\,$\mu$J. As the ionization potential of methanol is 10.84\,eV~\cite{NISTchemwebbook}, at least four photons of 3.10\,eV are required to ionize it. Table~\ref{tab:ioncount} lists the number of detected parent ions per second using an intensity of 20 or 100\,$\mu$J per pulse at a 1kHz repetition rate. Although the ion yield can probably be increased by further optimalization of the molecular beam parameters and focal properties of the laser beam, it seems unlikely that this changes the obtained results by more than a factor of 2. The fact that the nonresonant multi-photon ionization rate does not follow a $I^n$ dependence, with $I$ the laser intensity and $n$ the number of photons absorbed, can be understood by considering the intensity profile of the laser focus. At a certain laser intensity, $I_\text{sat}$, the ionization rate at the beam waist reaches unity, and the signal increases due to the increase in volume for which $I>I_\text{sat}$. For comparison, the ion yield for fluoromethane \ce{CH3F} is also listed in Table~\ref{tab:ioncount}. Note that the ionization potential of fluoromethane is 12.50\,eV~\cite{NISTchemwebbook} and at least five photons are required to ionize it.

\begin{table}[bth]
\centering
\caption{Number of methanol and fluoromethane parent ions resulting from a 400\,nm 120\,fs laser pulse at a repetition rate of 1\,kHz. For this measurement, methanol and fluoromethane molecules were seeded in argon with relative concentrations of 0.5 and 5\%, respectively.\label{tab:ioncount}}
\begingroup
\centering
\scriptsize
\begin{tabular*}{1 \columnwidth}{@{\extracolsep{\fill}}l c c c}

\hline\hline
\\[-2ex]
molecule      &  IE\footnote[1]{Data from Lias~\emph{et al.}~\cite{NISTchemwebbook}}(eV)&\multicolumn{2}{c}{$R$ (ions/sec)}   \\
\\[-2ex]\cline{3-4}\\[-2ex]
              &         & 20\,$\mu$J/pulse  &     100\,$\mu$J/pulse         \\
\\[-2ex]
\hline
\\[-2ex]
\ce{CH3OH} & 10.84 &  1200   & 20\,000           \\
\ce{CH3F}  & 12.50 &  3500   & 25\,000           \\
\\[-2ex]
\hline\hline

\end{tabular*}
\endgroup
\end{table}

\section{Estimated accuracy\label{sec:allandev}}

From the measured ion yield and the results from the simulations, we can now estimate the accuracy that can be obtained in the described beam machine. In general, the accuracy of a frequency measurement depends on the $Q$ factor of the system and the signal to noise ratio, $S/N$. In an experiment that relies on counting individual ions the accuracy is expressed by the Allan deviation~\cite{Vanier:Book}

\begin{equation}
\sigma_y(T)=\frac{1}{Q}\frac{1}{{S/N}}\sqrt{\frac{T_c}{T}}\quad\text{with }Q=\frac{\nu}{\Delta\nu},
\label{eq:allandev}
\end{equation}

\noindent
where $T_{c}$ defines the duration of one measurement cycle, $T$ is the total measurement time, $\nu$ is the frequency of the measured transition and $\Delta \nu$ the width of the spectral line.  In our simulations, we assume that the microwave cavity (or microwave cavities, if we use a Ramsey type setup) has a length of 160\,mm while the molecular beam has a velocity of 800\,m/s. This implies that the total interrogation time for the measurement is 200\,$\mu s$, corresponding to a spectral width of $\sim$4\,kHz. For the $3_{-1}\,E$ to 2$_{0}\,E$ transition at 12\,GHz, this results in a Q-factor of $3\times 10^{6}$. The $S/N$ depends on the number of ions that are detected per cycle. If we spend half the duration of a measurement cycle on and off resonance, the number of ions that contribute to the signal is given by $S=T_{c}|R_\text{off}-R_\text{on}|/2$. Assuming Poissonian statistics, the noise on the total number of detected ions is given by $N=\sqrt{T_{c}(R_\text{off}+R_\text{on})/2}$. Thus the Allan deviation becomes; 

\begin{equation}
\sigma_y(T)=\frac{1}{Q}\frac{\sqrt{R_\text{off}+R_\text{on}}}{|R_\text{off}-R_\text{on}|}\sqrt{\frac{2}{{T}}}
=\frac{1}{Q}\frac{\sqrt{\bar{R}}}{|\Delta R|}\frac{2}{\sqrt{R_0T}},
\label{eq:allandev2}
\end{equation}

\noindent
As expected, the Allan deviation becomes infinite when the count rates on and off resonance are equal, while it scales with $1/\sqrt{R_\text{off}}$ when $R_\text{on}$ is zero (background free). The extra factor of $\sqrt{2}$ arises because we spend half of the time on signal and half of the time off signal. The values for $|\Delta R|/\sqrt{\bar{R}}$ are given in Table~\ref{tab:transitions}. $R_{0}$, the count rate when the deflection fields are turned off, can be estimated from the test measurements. Assuming that the density in the molecular beam decreases quadratically with the distance from the nozzle, the expected signal at a distance of 1350\,mm behind the nozzle is about a factor of 64 smaller than that obtained in the test setup, hence $R_0=3.1\times 10^2$ ions/sec. For the $3_{-1}\,E$ to 2$_{0}\,E$ transition at 12\,GHz, the Allan deviation is $7.0\times 10^{-7}/\sqrt{T}$. In order to achieve a fractional accuracy of 10$^{-8}$ a measurement time of about 80 minutes would be required. Similar results are found for the other transition in \ce{CH3OH} listed in Table~\ref{tab:transitions}. The expected accuracy for \ce{CD3OH} is less favorable.

\section{Conclusion}

In this paper we estimate the feasibility of performing high resolution microwave spectroscopy on selected transitions in methanol using a Rabi-type molecular beam setup in combination with a femtosecond laser. We have adapted the {\sc belgi} programme to calculate the Stark effect for the different isotopologues of methanol. The calculated Stark shifts are reasonably large, typically between -1 and 1\,cm$^{-1}$ in a field of 100\,kV/cm for the different rotational states. Thus, the molecules can be easily manipulated using modest sized deflection fields. Due to the small rotational constants of methanol, many states are populated even at the low temperatures that can be obtained in supersonic beams. 
With the resolution obtained in a typical molecular beam deflection setup, it is not possible to select individual quantum states. Typically, 5 or more states are present within the laser focus. Furthermore, in the field free region used for inducing the microwave transition, the $M_{J}$ quantum number is not preserved, leading to a reduction of the state purity. Note that, $M_J$-scrambling is both a blessing and a cure, on one hand it reduces the observed difference between $R_\text{on}$ and $R_\text{off}$, on the other hand it ensures that all hyperfine levels are addressed without he need to change the position of the laser focus. $M_J$-scrambling can be enforced by rapidly switching the second deflection field on entrance of the molecules~\cite{vanVeldhoven2002}. From simulations, we find that the differences in signal on or off resonance, $\Delta R$, are in the range of $0.01-0.02$ for the selected transitions in \ce{CH3OH}, and slightly less for the two considered transitions in \ce{CD3OH}. In order to estimate the detection efficiency of methanol molecules using a femtosecond laser, we have performed test measurements in an existing molecular beam machine. Using a laser power of 100\,$\mu$J per pulse at 1\,kHz repetition rate, $2\,\times 10^4$ ions per second were detected at a distance of 160\,mm behind the nozzle. From these numbers, the described molecular beam deflection setup seems suited to measure microwave transitions in \ce{CH3OH} at an accuracy below 10$^{-8}$, although the required measurement times are rather long. Currently the laser system is being upgraded to provide peak intensities that are 5 times higher than used in this study. This should result in an increase of the ion yield of at least a factor of 25, making this experiment perfectly feasible. Due to the small differences in count rate, however, it might be tedious to find the optimal position of the laser focus. For a laboratory test using \ce{CD3OH} an accuracy well below 10$^{-10}$ would be required. This seems unlikely to be achieved in a beam machine as considered here.

\section*{Acknowledgements}
This research has been supported by FOM via a Projectruimte grant. R.M.L. acknowledges financial support from NSERC Canada. W.U. acknowledges financial support from the Templeton Foundation. H.L.B acknowledges financial support from NWO via a VIDI grant and from the ERC via a Starting grant. We thank C.S. Lehmann for helpful discussions and assistance during the experiment.

%\bibliographystyle{tMPH}
%\bibliography{jansen_et_al}

 \newcommand{\noopsort}[1]{} \newcommand{\printfirst}[2]{#1}
  \newcommand{\singleletter}[1]{#1} \newcommand{\switchargs}[2]{#2#1}

\end{document}